\documentclass[reprint,amsmath,amssymb,apr,aip,twocolumn]{revtex4-1}
\usepackage{graphicx}
\usepackage{graphics}
\usepackage{algorithm,algorithmic}
\usepackage{mathptmx}
\usepackage{times}
\usepackage{amsmath}
\usepackage{amssymb}
\usepackage{dcolumn}
\usepackage{color}
\usepackage{bm}
\usepackage{units}

\begin{document}
\title{Mixed-Precision In-Memory Computing}
\author{Manuel Le Gallo*}
%\email{anu@zurich.ibm.com}
\affiliation{IBM Research - Zurich, 8803 R\"{u}schlikon, Switzerland}
\affiliation{ETH Zurich, 8092 Zurich, Switzerland}
\author{Abu Sebastian*}
%\email{ase@zurich.ibm.com}
\affiliation{IBM Research - Zurich, 8803 R\"{u}schlikon, Switzerland}
\author{Roland Mathis}
\affiliation{IBM Research - Zurich, 8803 R\"{u}schlikon, Switzerland}
\author{Matteo Manica}
\affiliation{IBM Research - Zurich, 8803 R\"{u}schlikon, Switzerland}
\affiliation{ETH Zurich, 8092 Zurich, Switzerland}
\author{Heiner Giefers}
\affiliation{IBM Research - Zurich, 8803 R\"{u}schlikon, Switzerland}
\author{Tomas Tuma}
\affiliation{IBM Research - Zurich, 8803 R\"{u}schlikon, Switzerland}
\author{Costas Bekas}
\affiliation{IBM Research - Zurich, 8803 R\"{u}schlikon, Switzerland}
\author{Alessandro Curioni}
\affiliation{IBM Research - Zurich, 8803 R\"{u}schlikon, Switzerland}
\author{Evangelos Eleftheriou}
\affiliation{IBM Research - Zurich, 8803 R\"{u}schlikon, Switzerland}
\date{\today}
%----------------------------------------------------------------------------------------------------------------------------------------------------
\begin{abstract}

As CMOS scaling reaches its technological limits, a radical departure from traditional von Neumann systems, which involve separate processing and memory units, is needed in order to significantly extend the performance of today's computers. In-memory computing is a promising approach in which nanoscale resistive memory devices, organized in a computational memory unit, are used for both processing and memory. However, to reach the numerical accuracy typically required for data analytics and scientific computing, limitations arising from device variability and non-ideal device characteristics need to be addressed. Here we introduce the concept of mixed-precision in-memory computing, which combines a von Neumann machine with a computational memory unit. In this hybrid system, the computational memory unit performs the bulk of a computational task, while the von Neumann machine implements a backward method to iteratively improve the accuracy of the solution. The system therefore benefits from both the high precision of digital computing and the energy/areal efficiency of in-memory computing. We experimentally demonstrate the efficacy of the approach by accurately solving systems of linear equations, in particular, a system of $5,000$ equations using $998,752$ phase-change memory devices.

\end{abstract}
%--------------------------------------------------------------------------------------------------------------------------------------------------------
\maketitle

Nanoscale resistive memory devices, which are also referred to as memristive devices, can store information in their conductance states and can remember the history of the current that has flowed through them \cite{Y2008strukovNature,Y2011chuaAPA,Y2015wongNatureNano}. These devices form the basis of in-memory computing: an approach in which both information processing and storing computational data are performed on the same physical devices organized in a computational memory unit \cite{Y2013diventraNatPhys,Y2015traversaTNNLS,sebastian2017,LeGallo2017}. With such systems, various physical mechanisms, including Ohm's law and Kirchhoff's circuit laws\cite{hu2016,li2018}, chemically-driven phase transformations\cite{Y2013xuSR}, the pattern dynamics of ferroelectric domain switching\cite{Y2014ievlevNatPhys}, and the physics of crystallization\cite{Y2013ielminiAdvMat,Y2014sebastianNatComm} and melting\cite{Y2014lokePNAS} in phase-change materials, can be used to perform arithmetic\cite{Y2011wrightAdvMat,Y2013xuSR,Y2015hosseiniEDL,hu2016} and logical\cite{Y2010borghettiNature,Y2014kvatinskyTCAS,Y2013ielminiAdvMat,Y2014lokePNAS} operations. Research on these devices has already led to the development of massively parallel, memory-centric hardware accelerators with applications ranging from image processing to healthcare\cite{2016bojnordiHPCA,shafiee2016,sheridan2017,Y2015choiSR}. However, building a computational memory unit that can solve practical problems in a reliable and accurate way remains challenging. Memristive devices suffer from significant inter-device variability and inhomogeneity across an array\cite{Y2014ambrogioITED}. Moreover, they exhibit intra-device variability and a randomness that is intrinsic to the way the devices operate\cite{Y2013fantiniIMW,Y2016legalloESSDERC}. While this randomness can be exploited for certain types of computational tasks\cite{Y2013gabaNanoscale,Y2016tumaNatureNano}, the low precision associated with computational memory is prohibitive for many practical applications.

\begin{figure*}[t!]
\centering
\begin{tabular}{c}
\includegraphics[width = 2\columnwidth]{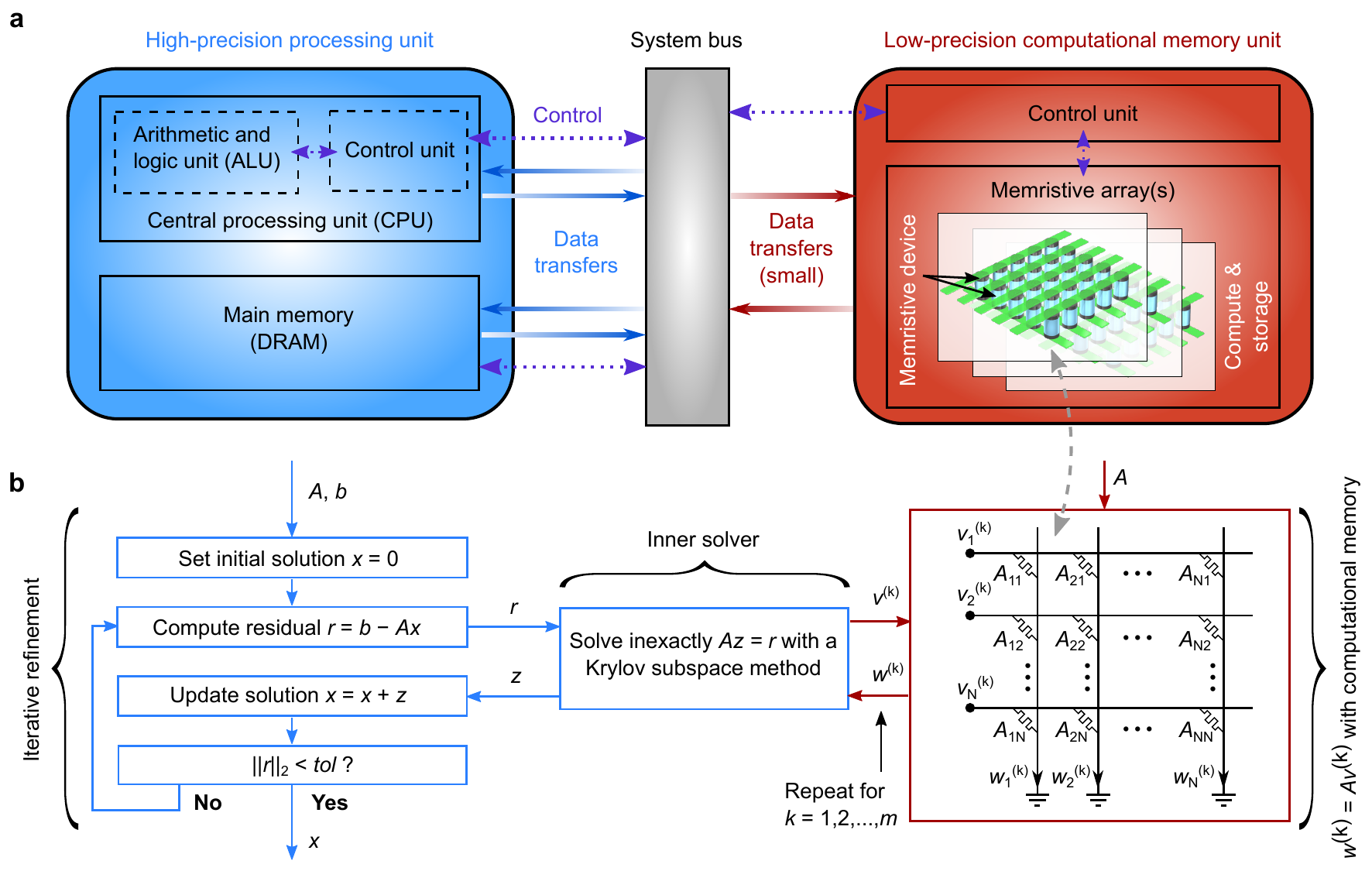}
\end{tabular}
\caption{\textbf{Concept of mixed-precision in-memory computing. } \textbf{a}, Possible architecture of a mixed-precision in-memory computing system. The
high-precision processing unit (left) performs digital logic computation and is based on the standard von Neumann computing architecture. The
low-precision computational memory unit (right) performs analog in-memory computation using one or multiple memristive arrays. The system bus (middle)
implements the overall management (control, data, addressing) between the two units. The purple dotted arrows indicate control communication and the
solid arrows (red, blue) indicate data transfers. \textbf{b}, Algorithm for solving a system of linear equations $Ax=b$ using the mixed-precision in-memory computing
system of \textbf{a}. The blue boxes show the steps implemented in the high-precision processing unit, and the red box shows the matrix-vector
multiplication step implemented in the low-precision computational memory unit. } \label{fig:1}
\end{figure*}

In this article, we introduce the concept of mixed-precision in-memory computing to address this issue. The concept is motivated by the observation that many computational tasks can be formulated as a sequence of two distinct parts. In the first part, an approximate solution is obtained. In the second part, the resulting error in the overall objective is calculated accurately. Then, based on this, the approximate solution is adapted (by repeating the first part). The first part typically has a high computational load, whereas the second part has a light computational load. By repeating this sequence several times, it is often possible to arrive at a solution with arbitrarily high accuracy\cite{bekas2009}. In a mixed-precision in-memory computing system, the idea is to use a low-precision computational memory unit to obtain the approximate solution of the first part and a high-precision processing unit to realize the second part (Fig.\ \ref{fig:1}a). The expectation is that in this way we can benefit from an overall high areal and energy efficiency, because the bulk of the computation is still realized in a non-von Neumann manner, and still be able to achieve an arbitrarily high computational accuracy.

\section{Mixed-precision in-memory linear equation solver}

%--------------------------------------------------------------------------------------------------------------------------------------------
% Linear equation solver
% Why is it interesting
% Iterative solvers: Krylov subspace methods
% Mixed precision approach with matrix-vector multiplication in a memcomputer
%--------------------------------------------------------------------------------------------------------------------------------------------
To illustrate this concept, we present the problem of solving systems of linear equations. The problem is to find an unknown vector $x \in
\mathbb{R}^{N}$ that satisfies the constraint
\begin{equation}\label{eqn:lin}
Ax=b, \text{ where } A \in \mathbb{R}^{N \times N} \text{ and }b \in \mathbb{R}^{N}.
\end{equation}
Here $A$ is a non-singular matrix and $b$ is a known column vector of $N$ observations or measurements. The target of our study is the solution of dense covariance matrix problems, which are common in cognitive computing and data analytics.\cite{bekas2009}
%In industrially relevant applications such as xxx $N$ can be on the order of x and solving \eqref{eqn:lin} demands extensive computational resources (\textbf{check with Costas}). 
Such problems can be solved in the mixed-precision in-memory computing framework as shown in Fig. \ref{fig:1}b. In a so-called
iterative refinement algorithm, an initial solution is chosen as starting point and is iteratively updated with a low-precision error-correction
term, $z$. The error-correction term is computed by solving $Az=r$ with an inexact inner solver using the residual $r = b-Ax$, calculated with high
precision.\cite{klavik2014} The algorithm runs until the norm of the residual falls below a desired tolerance, $tol$. 

For the inner solver, we use an iterative Krylov subspace method, such as the Conjugate Gradient method or the Generalized Minimum Residual (GMRES) method.\cite{saad2003} Krylov subspace methods are currently considered to be among the most important iterative techniques available for solving \eqref{eqn:lin} with high dimensional matrices \cite{saad2003}. These techniques rely on building a basis $\lbrace v^{(k)} \rbrace_{k=1}^m$ of the $m$-th Krylov subspace $\mathcal{K}_m(A,r) = \mathrm{span}\lbrace r,Ar,A^2r,...,A^{m-1}r \rbrace$. This basis is obtained by performing multiple matrix-vector multiplications $w^{(k)} = Av^{(k)}$ with the matrix $A$, and using $w^{(k)}$ to compute the next basis vector $v^{(k+1)}$ following an orthogonalization procedure. From this basis, the error-correction term, which is an approximation of $A^{-1}r$, can be obtained. 

In all Krylov subspace methods, the computationally most intensive operation is the matrix-vector multiplication $w^{(k)} = Av^{(k)}$. Hence, the key idea is to realize this operation in the computational memory unit, using a memristive crossbar array in which matrix $A$ is programmed as the conductance values of the memristive devices (Fig. \ref{fig:1}b). 
%If a voltage $V$ is applied to a device with conductance $G$, the resulting current $I$ is the result %of the $GV$ multiplication if Ohm's law is applicable. By appropriately summing the currents from the multiple memristive devices, the matrix-vector computation is effectively performed.
This mode of computing is very efficient because the matrix-vector product is computed \textit{in situ} in the memristive array, thereby eliminating any intermediate movement of data.\cite{hu2016} Even if the computation realized this way is approximate and introduces perturbations in the inner solver, the outer iterative refinement algorithm ensures convergence to a high-accuracy solution.\cite{klavik2014}  The magnitude of the perturbations that can be tolerated is expected to decrease with increasing condition number of matrix $A$ (the condition number reflects how much the solution $x$ will change with respect to a change in $b$).\cite{higham2002}
%Therefore, if the problem \eqref{eqn:lin} is not too ill-conditioned, it is possible to realize the matrix-vector computations of the inner solver
%using low-precision memcomputing and expect the iterative refinement algorithm to produce a solution with the desired high accuracy.

\section{In-memory multiplications with PCM devices}
%--------------------------------------------------------------------------------------------------------------------------------------------
% Experimental results
% Platform, overview
% Scalar multiplication
% Matrix-vector multiplication
%--------------------------------------------------------------------------------------------------------------------------------------------

\begin{figure*}[t!]
\centering
\begin{tabular}{c}
\includegraphics[width = 2\columnwidth]{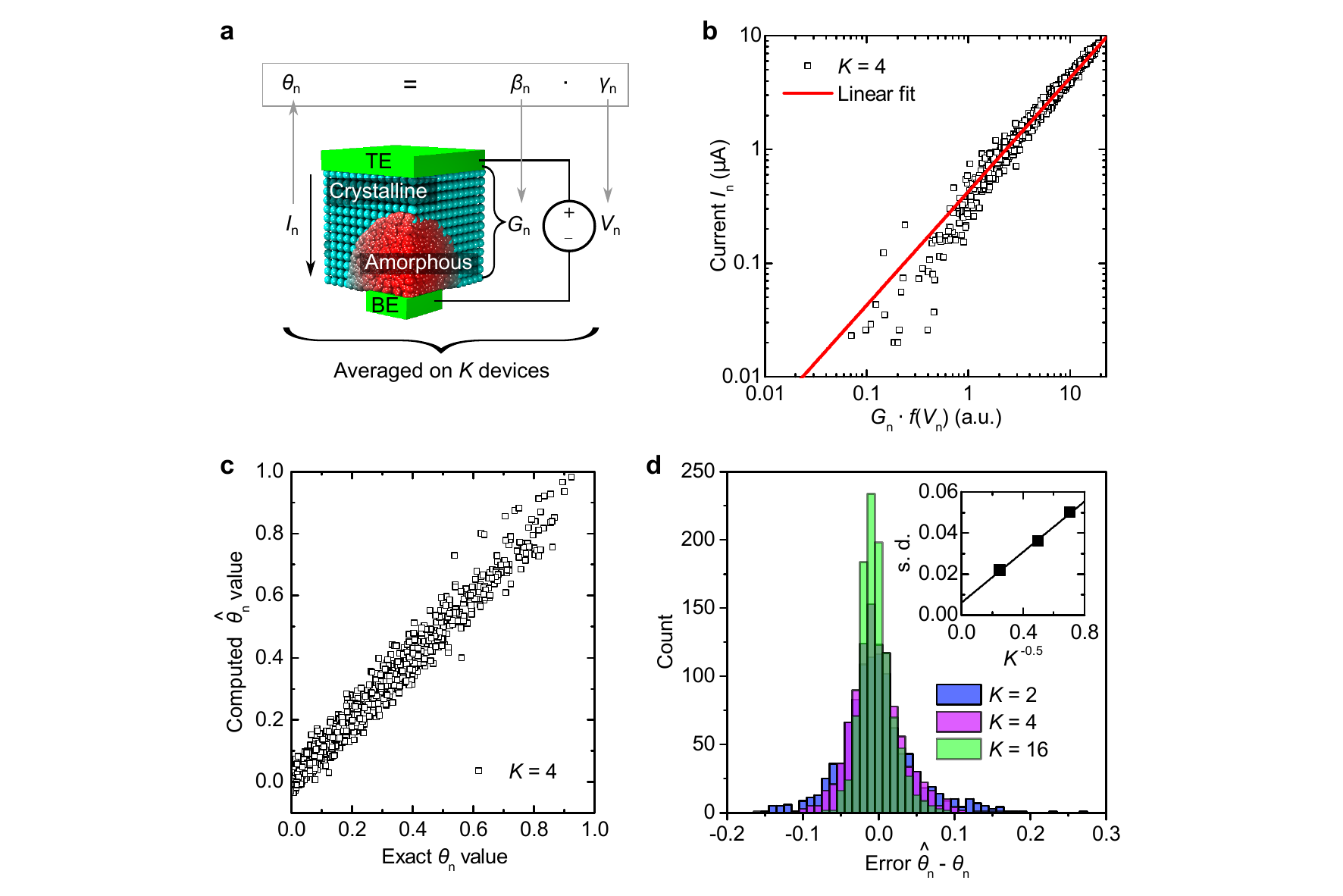}
\end{tabular}
\caption{\textbf{Scalar multiplication. } \textbf{a}, Schematic of a PCM device and the scalar multiplication implementation based on Ohm's
law. TE (BE) denotes top (bottom) electrode. The grey arrows indicate mappings from one variable to another. \textbf{b}, Plot showing the proportionality
between $I_n$ and $G_n f(V_n)$ (Eq. \eqref{eqn:pseudoohm}) for the 1024 different combinations of $\lbrace \beta_n,\gamma_n \rbrace$. \textbf{c},
Final result of the computed scalar multiplication $\hat{\theta}_n$ plotted against the exact result $\theta_n$. \textbf{d}, Error distributions for different numbers of averaged devices $K$. The inset in \textbf{d} shows the standard deviation (s.d.) of the distributions versus $K^{-0.5}$. }
\label{fig:2}
\end{figure*}

For our experiments, we implemented the low-precision matrix-vector multiplication using a prototype chip containing one million phase-change
memory (PCM) devices. PCM devices are resistive memory devices that can be programmed to achieve a desired conductance value by altering the
amorphous/crystalline phase configuration within the device (Fig. \ref{fig:2}a).\cite{Y2016burrJETCAS} The array consists of a matrix of 512 word
lines $\times$ 2048 bit lines integrated in 90-nm CMOS technology and connected in a crossbar configuration. Each crosspoint consists of a PCM device in series with an access transistor (see Methods and Supplementary Note I). 

First, we investigated the scalar multiplication operation that forms the core of the
matrix-vector multiplication performed with the PCM devices. Let $\theta_n=\beta_n \cdot \gamma_n$, where $\beta_n$ and $\gamma_n$ are numbers generated
uniformly in $[0,1]$.  $\beta_n$ was mapped to an effective conductance value $G_n$ ($I/V$ ratio at \unit[$V=0.2$]{V}) between approximately 0 and
\unit[50]{$\mu$S}, and $\gamma_n$ to a voltage $V_n$ between approximately \unit[0.1]{V} and \unit[0.3]{V} (see Supplementary Note II). Because the
current is a slightly nonlinear function of the voltage in our PCM devices, the analogue multiplication was assumed to follow a ``pseudo'' Ohm's
law:
\begin{equation}\label{eqn:pseudoohm}
I_n \simeq \alpha G_n f(V_n).
\end{equation}
In this equation, $\alpha$ is an adjustable parameter and $f$ a polynomial function that approximates the current/voltage characteristics of the
PCM devices (Supplementary Note II). The devices were programmed to the effective conductance $G_n$ using an iterative program-and-verify procedure (see Methods)
and were subsequently read by applying a voltage $V_n$. The experiment was repeated for $n=1,\ldots,1024$ different combinations of $\lbrace \beta_n,\gamma_n
\rbrace$, and the results for each value of $n$ were averaged on $K$ devices (thus using $1024 \times K$ devices in total). As shown in Fig.
\ref{fig:2}b, the computation of Eq. \eqref{eqn:pseudoohm} is effectively realized over approximately 2 decades of current. The current, $I_n$, can
then be converted to an approximate value $\hat{\theta}_n$ that represents the final result of the computation (Supplementary Note II), which is plotted
in Fig. \ref{fig:2}c against the exact result $\theta_n$ computed in double-precision floating point. The distributions of the error
$\hat{\theta}_n-\theta_n$ get narrower with increasing $K$ (see Fig. \ref{fig:2}d), with the standard deviation scaling as $K^{-0.5}$ (see inset) as
dictated by the central limit theorem when averaging independent and identically distributed (iid) random variables. This indicates that the predominant
part of the error comes from random perturbations in the current $I_n$. Possible causes for such perturbations are inaccuracies in the iterative programming of the devices to the conductance $G_n$, variability of the current/voltage characteristics across devices, inherent conductance variations and low-frequency noise arising from the amorphous phase-change material \cite{Y2015koelmansNatComm}.

The matrix-vector multiplication is a natural extension of the scalar multiplication in which the elements of matrix $A$ are coded into the conductance
states of PCM devices. Because our experimental hardware only allows serial access to each individual crosspoint, only the element-by-element
multiplications of the matrix-vector product were performed in hardware, whereas the sum was performed outside of the chip (Supplementary Note III). The
accumulated effect of errors in this mode of computing is fundamentally different from that of rounding errors arising for example from fixed-point
data conversions \cite{higham2002} (Supplementary Note IV). 
To prevent errors in the multiplication results due to
the temporal evolution of the conductance values (drift)\cite{SebastianIRPS2015,LeGalloIRPS2016} of the PCM devices, we developed a
calibration procedure which consists of periodically reading the
summed conductance of a subset of the devices encoding matrix $A$ to account for a
global conductance shift during an experiment (Supplementary Note V). This
simple procedure is easily implemented in a crossbar array, and
estimates the conductance variations directly from the devices
without any assumptions on how the conductance changes. 

%The structural relaxation of the amorphous phase to an energetically more favorable
%``ideal glass" state and its manifestation as a temporal evolution of the conductance values (drift) also poses challenges that need to be accounted for
%(Supplementary Note V).

\section{Accurately solving linear equations using PCM hardware}

%--------------------------------------------------------------------------------------------------------------------------------------------
% Model matrix
%--------------------------------------------------------------------------------------------------------------------------------------------

\begin{figure}[t!]
\centering
\begin{tabular}{c}
\includegraphics[width = \columnwidth]{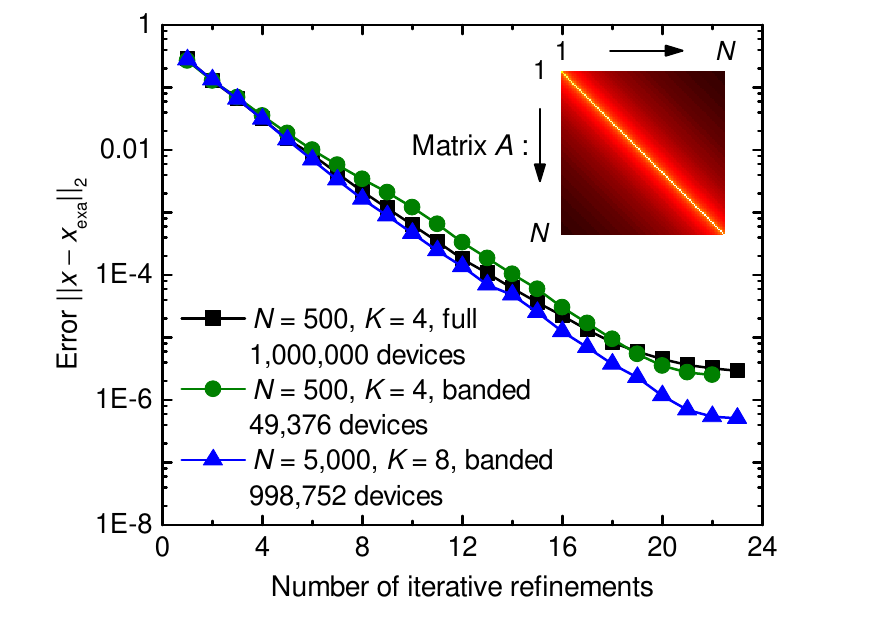}
\end{tabular}
\caption{\textbf{Solution of a system of linear equations involving a model covariance matrix. } Norm of error between the experimentally obtained $x$ and the exact solution
$x_{\mathrm{exa}}$ of Eq. \eqref{eqn:lin} as a function of the number of iterative refinements. $x_{\mathrm{exa}}$ was
computed by direct inversion of Eq. \eqref{eqn:lin} in double-precision floating point. The norm of error converges to a value that is determined by the system of linear equations and the desired tolerance for the norm of the residual, $tol$. $m=5$ inner solver iterations were used for $N=500$ and $m=10$ for $N=5,000$. $K$ denotes the number of PCM devices averaged to represent one matrix
element. ``Full'' means that the full matrix $A$ is programmed in the PCM chip, and ``banded'' means that a reduced banded version of $A$ with $12$ entries on each side of the main diagonal is programmed in the PCM chip. The inset shows a heat map (colormap in log scale) of the model covariance matrix $A$ for $N=500$. } \label{fig:3}
\end{figure}

Next, we present the solution of \eqref{eqn:lin} for model covariance matrices of different sizes defined as
\begin{equation}\label{eqn:modelmatrix}
A_{ij}=
\begin{cases}
|i-j|^{-1} ,& \text{if } i \neq j \\
1+\sqrt{i},& \text{if } i = j
\end{cases}
\end{equation}
%$A_{ij}^{i \neq j} = |i-j|^{-1}, A_{ij}^{i=j} = 1+\sqrt{i}$ 
for $i=1,...,N$ and $j=1,...,N$. Such matrices exhibit a decaying behavior that simulates decreasing correlation of features away from the main diagonal.\cite{bekas2009} 
%The elements of $b$ were generated uniformly in $[0,1]$. 
%The inner solver was chosen to be Conjugate Gradient (Supplementary Note VI). 
We first programmed a full matrix \eqref{eqn:modelmatrix} of size $N=500$ in our PCM chip with $K=4$ devices averaged per matrix element, using all one million PCM devices available, and executed the mixed-precision algorithm with Conjugate Gradient (CG) as inner solver for $tol=10^{-5}$ (see Methods). The experiment converged to the desired accuracy after 23 iterative refinements (see Fig. \ref{fig:3}). 

In the mixed-precision computing framework, one can work not only with an imprecise inner solver, but also with inexact input data.\cite{klavik2014} For instance, because the elements far from the main diagonal of \eqref{eqn:modelmatrix} are small, a reduced banded version of the matrix (with just $12$ entries on each side of the main diagonal) can be coded in the memristive array instead of the complete one. In this way, the inner solver works on an \textit{inexact} version of matrix $A$, which is coded in the memristive array, whereas the outer iterative refinement loop works towards finding the \textit{exact} solution of \eqref{eqn:lin} by using the full matrix $A$ for computing the residuals. Using this approach with matrix size $N=500$ and $K=4$, we obtained a convergence rate almost identical to that without banding (see Fig. \ref{fig:3}). We then tested this approach up to the maximum matrix size for which we could program the banded matrix in the PCM chip, and obtained the desired convergence for $N=5,000$ using $K=8$ (see Fig. \ref{fig:3}). 
23 high-precision matrix-vector multiplications were required to solve this problem with mixed-precision in-memory computing, whereas 50 matrix-vector multiplications are needed when performing a single run of the CG algorithm in high precision to obtain the same solution accuracy. Therefore, the mixed-precision in-memory computing solution indeed reduces the number of floating point operations and associated data transfers needed to solve the problem compared to a conventional von Neumann implementation in high precision. 

Because of the high-precision iterative refinement, the maximum achievable accuracy of the mixed-precision in-memory computing system is limited only by the precision of the high-precision processing unit, but not by the precision of the computational memory unit. The minimum error reached experimentally for $N=500$ when setting $tol=10^{-15}$ is $\sim 1.3 \cdot 10^{-15}$, which is limited by the machine precision of the high-precision processing unit we use (Supplementary Note VI). Several methods can be used to speed up the convergence of the mixed-precision algorithm further and allowed us to obtain convergence for even larger matrix sizes of up to $N=10,000$ (Supplementary Note VI).

\begin{figure*}[t!]
\centering
\begin{tabular}{c}
\includegraphics[width = 2\columnwidth]{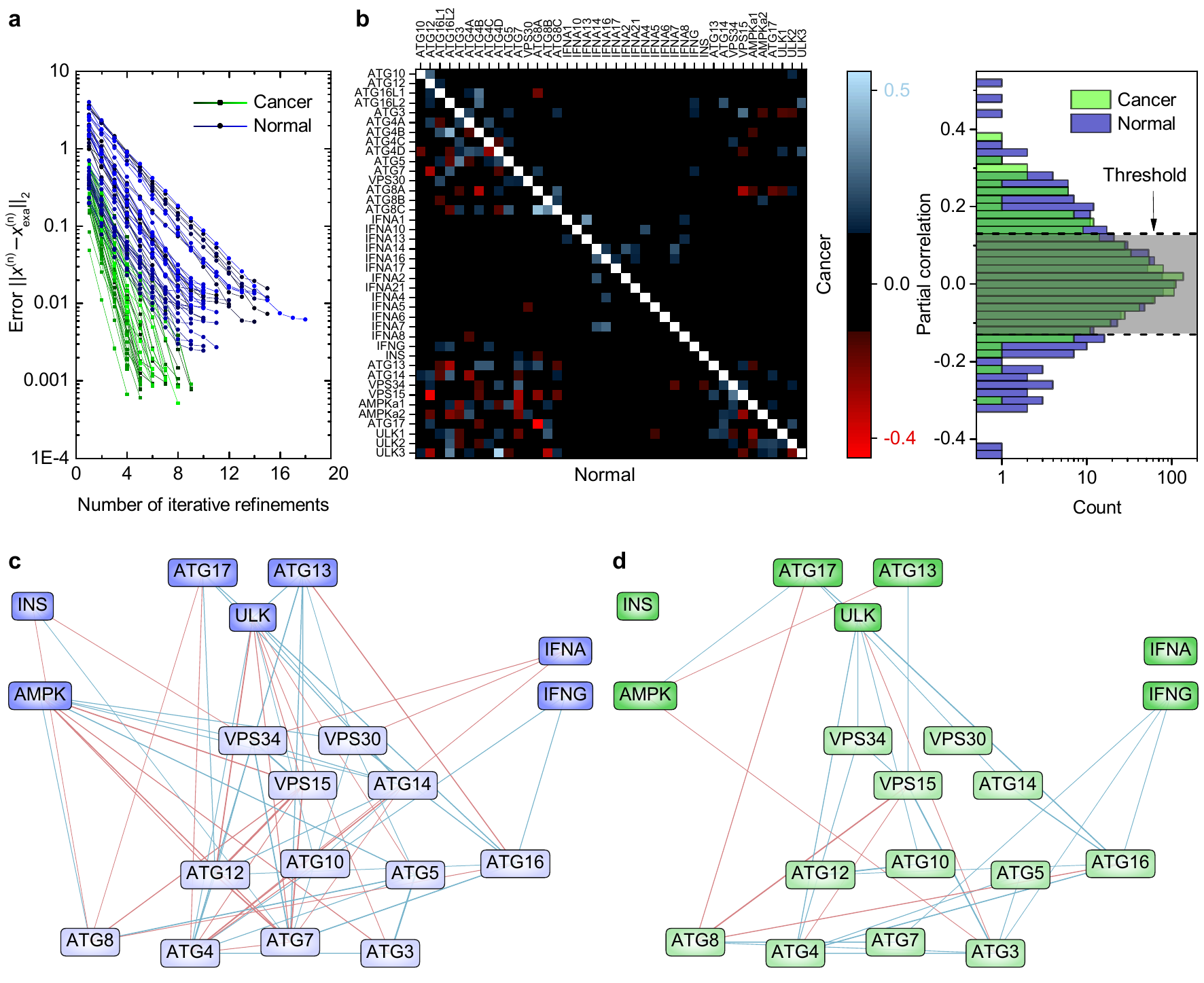}
\end{tabular}
\caption{\textbf{Estimation of autophagy-related gene interactions from RNA measurements. } \textbf{a}, Experimentally obtained convergence of the mixed-precision algorithm for the 40 linear equations solved for the cancer and the normal tissues. $x^{(n)}_{\mathrm{exa}}$ was computed by direct inversion of Eq. \eqref{eqn:lin} in
double-precision floating point. $m=5$ inner solver iterations were used. \textbf{b}, Matrix of computed partial correlations of the 40 genes studied for cancer and normal tissues (left)
and their distributions (right). For visualization purposes, only the interactions for which the magnitude of the partial correlations is
larger than a threshold of 0.13, corresponding to the 90-th percentile of the normal tissue, are displayed. \textbf{c}, Interactome obtained from normal
tissue. \textbf{d}, Interactome obtained from cancer tissue. In \textbf{c} and \textbf{d}, the upstream nodes are dark colored and the downstream
targets are light colored. Blue edges denote positive interactions and red edges denote negative interactions.} \label{fig:4}
\end{figure*}

In addition, we tested the mixed-precision algorithm on a practical problem for which matrix $A$ was built from real-world data.
For this, we used RNA expression measurements of genes obtained from cancer patients, publicly available from The Cancer Genome Atlas (TCGA) project
(see Methods). We focused our investigation on 40 genes reported in the manually curated autophagy pathway of the Kyoto Encyclopedia
of Genes and Genomes (KEGG). 
%Autophagy has been associated with two opposing roles in cancer. It acts as a tumor suppressor by degrading damaged proteins and organelles, and it enables tumors to tolerate metabolic stress \cite{Y2007mathewNatRevCancer,Y2011yangMCT}. 
Autophagy plays opposing roles in cancer by both acting as a tumor suppressor by degrading damaged proteins and organelles as well as enabling tumors to tolerate metabolic stress \cite{Y2007mathewNatRevCancer,Y2011yangMCT}. 
To infer and compare the
networks of gene interactions (interactomes) from normal and cancer tissues, we calculated the partial correlations between the genes by computing
the inverse covariance matrix $\Sigma$ from 946 normal tissue samples and from 946 cancer tissue samples (see Methods). Given
the covariance matrix $A$ of the 40 genes, $\Sigma$ can be obtained by solving $Ax^{(n)}=e^{(n)}$ for $n=1,...,40$, where $e^{(n)}$ has all entries equal to zero
except the $n$-th one, which is 1, and $x^{(n)}$ is the resulting $n$-th column of $\Sigma$. 

We programmed the $40\times40$ covariance matrix in the
PCM chip and used mixed-precision in-memory computing with GMRES as inner solver to solve the 40 linear equations (see Methods). The procedure was repeated for both cancer and normal tissues.
The algorithm converged to the desired precision for all 40 linear systems solved (see Fig. \ref{fig:4}a) and the resulting $\Sigma$ matrix was
sufficiently accurate for computing the interactome (the interactomes obtained with the exact and computed $\Sigma$ are identical). The computed
partial correlations of the 40 genes studied and their distributions are shown in Fig. \ref{fig:4}b. While some of the gene interactions are preserved between the cancer and normal tissues, the cancer network exhibits a different connectivity pattern (see Fig. \ref{fig:4}c and \ref{fig:4}d). 
%The cancer network is less connected than the network inferred from normal tissue. 
In the normal tissue, the upstream signals INS, AMPK, ULK, ATG13,
ATG17, IFNA and IFNG (dark colored) correlate with many of the downstream targets (light colored) known to be involved in the formation of
autophagosomes, the molecular agents of autophagy. The partial correlations computed on cancerous tissue yield a sparsely connected network, implying
an altered regulation pattern, as is commonly observed in cancer \cite{Y2012westSR,Y2010schrammBMC,Y2013hongNAR}. 
%Such analysis could be scaled up to larger matrices and combined with stochastic estimators for computing the inverse covariance, which require solving only $s \ll N$ linear systems. The methods used here thus have the potential to provide much-needed energy-efficient solutions to analyze the ever-increasing amounts of biological data. 
%More details can be found in Supplementary Note VII.

\section{Performance assessment and current limitations}
%--------------------------------------------------------------------------------------------------------------------------------------------
% Significance of mixed-precision in-memory computing
%--------------------------------------------------------------------------------------------------------------------------------------------
The above demonstration highlights the importance of linear analysis in problems associated with cognitive computing and data analytics. The fact that such computation can be performed partly with computational memory without sacrificing the overall computational accuracy opens up exciting new avenues towards energy-efficient large-scale data analytics, in which the massive data transfers inherent to the traditional von Neumann architecture have become the most energy-hungry part. Such solutions are much needed because analyzing the ever-growing datasets we produce will quickly increase the computational load to the exascale level if standard techniques are to be used.\cite{bekas2009} 

%--------------------------------------------------------------------------------------------------------------------------------------------
% Applicability and shortcomings of current hardware
%--------------------------------------------------------------------------------------------------------------------------------------------

The problems tackled in this work were well-conditioned and of relatively small scale because of the limited size and
precision of our hardware. Scale-up strategies include building larger arrays and/or operating several of them in parallel. To address problems with a broader range of condition numbers, it will be necessary to increase the precision of the computational memory unit beyond that achieved in the present work to allow the Krylov-subspace inner solver to converge more easily \cite{klavik2014}. Possible avenues are improving the memristive device characteristics with respect to variability and conductance noise \cite{Y2015koelmansNatComm}, mapping a single column of the matrix to multiple physical columns of an array encoding different bits (Supplementary Note III), and using error-correction techniques within the computational memory unit \cite{feinberg2018}. The efficiency and robustness of iterative Krylov-subspace solvers can also be improved by using preconditioning techniques \cite{saad2003}. 
Moreover, although we restricted our experiments to diagonally-dominant covariance type matrices, it is not a limitation of the mixed-precision in-memory computing concept, which can be used to solve \eqref{eqn:lin} for more general types of matrices provided that the convergence conditions\cite{saad2003} of the chosen Krylov subspace solver are met. In fact, while the CG method requires that the matrix $A$ is symmetric and positive-definite, the GMRES method that we used on the RNA data can deal with a much broader family of problems, in particular non-symmetric matrices, but at a slightly higher computational complexity than CG \cite{saad2003}. 

%--------------------------------------------------------------------------------------------------------------------------------------------
% Performance assessment
%--------------------------------------------------------------------------------------------------------------------------------------------
Finally, we performed a detailed study to compare the energy efficiency of the mixed-precision in-memory computing system with that of conventional von Neumann implementations (Supplementary Note VII.A). We implemented all data conversions and data transfers between the high-precision and computational memory units as well as all additional floating-point operations needed to solve \eqref{eqn:lin} with the algorithm of Fig. \ref{fig:1}b. We then experimentally measured the runtime and power consumption of the system using both a IBM POWER8 CPU and a NVIDIA P100 GPU as high-precision processing unit. In those measurements, it was assumed that the operations in the computational memory unit consume negligible time and energy compared to the operations performed in the high-precision computing unit, thus providing an upper bound on the achievable performance of the system. Our analysis shows that mixed-precision in-memory computing can outperform both CPU-based and GPU-based implementations that use only high-precision arithmetic in terms of both time and energy to solution. For matrix \eqref{eqn:modelmatrix}, the maximum measured dynamic energy gains range from $6.8\times$ with the precision of the computational memory unit comparable to that of our current PCM chip, up to $24\times$ when assuming two orders of magnitude less noise\cite{Y2015koelmansNatComm} in the computational memory unit. Note that those numbers are strongly tied to matrix $A$ and the right-hand side $b$ used, and that in general higher energy gains are expected the more ill-conditioned the matrix $A$ is \cite{anzt2012}. The achievable performance depends on the ratio between the number of iterations required in the high-precision-only implementation versus the number of iterative refinements performed in the mixed-precision in-memory computing algorithm, which in turn depends on both the precision of the computational memory unit and the actual problem that is solved (Supplementary Note VII.A).

Subsequently, we derived specifications which should be met by the computational memory unit in order to achieve a system performance close to the aforementioned upper bound in terms of speed. Assuming a crossbar size of $1000 \times 1000$ cells, we expect that operating 10 crossbars in parallel at a cycle time of \unit[1]{$\mu$s} or less should allow the mixed-precision in-memory computing system to reach optimal performance in terms of speed (Supplementary Note VII.B). We believe that those specifications should be within the reach of existing technology because circuit simulations show that memristive crossbars can be run at a frequency of \unit[10]{MHz} \cite{hu2016} and $128 \times 64$ memristive crossbars with \unit[$<100$]{ns} latency have already been demonstrated \cite{li2018}. During the execution of the mixed-precision algorithm, only read operations are performed on the memristive array, which consume much less energy than programming (\unit[$1-100$]{fJ} per PCM device), and hence the additional power overhead from the crossbar is expected to be minimal compared to that of the high-precision processing unit (Supplementary Note VII.B). Nonetheless, efficient designs of the crossbar peripheral circuitry and I/O converters will be of utmost importance to ensure that the computational memory unit meets those specifications. 

Moreover, to assess the capability of computational memory to compete with already existing low-precision CMOS-based accelerators for performing matrix-vector multiplications, we designed a low-precision 4-bit matrix-vector multiplier on a field-programmable gate array (FPGA) with an accuracy comparable to what could be obtained with our current prototype PCM chip. Our analysis shows that even when all matrix coefficients are stored in the FPGA memory (thus neglecting any off-chip data transfers), a memristive crossbar array based on devices similar to our prototype PCM chip for performing analogue matrix-vector multiplications could already offer up to 80 times lower energy consumption than the FPGA solution (Supplementary Note VIII).

\section{Conclusions}
In summary, we have introduced the concept of mixed-precision in-memory computing to address the inherent imprecision associated with computational memory. The hybrid system comprises a computational memory unit, which performs the bulk of a given computational task, and a high-precision processing unit which implements a backward method to iteratively improve the accuracy of the solution. In this way, it is possible to achieve an arbitrarily high solution accuracy with the bulk of the computation realized as low-precision in-memory computing. We have experimentally demonstrated this concept by solving systems of linear equations using a PCM chip to perform analogue matrix-vector multiplications, on both model covariance matrices and a practical problem in which the matrix was built from real-world RNA expression data. The next steps will be to generalize mixed-precision in-memory computing beyond the application domain of solving systems of linear equations to other computationally intensive tasks arising in automatic control, optimization problems, machine learning, deep learning\cite{nandakumar2017}, and signal processing.
%We finally stress that mixed-precision memcomputing can be used in applications that extend beyond the solution of linear equations to
%other relevant computational tasks arising in automatic control, optimization problems, machine learning and signal processing.
\newpage

\section*{Methods}

\subsection*{Experimental platform.}

The experimental platform is built around a prototype PCM chip that comprises 3 million PCM devices. The PCM array is organized as a matrix of word lines (WL) and bit lines (BL). In addition to the PCM devices, the prototype chip integrates the circuitry for device addressing and for write and read operations. The PCM chip is interfaced to a hardware platform comprising two field programmable gate array (FPGA) boards and an analog-front-end (AFE) board. The AFE board provides the power supplies as well as the voltage and current reference sources for the PCM chip. The FPGA boards are used to implement overall system control and data management as well as the interface with the data
processing unit. The experimental platform is operated from a host computer, and a Matlab environment is used to coordinate the experiments. The algorithms used to solve the linear equations and all data conversions are implemented in Matlab software. 

The PCM devices were integrated into the chip in 90-nm CMOS technology
using the key-hole process described in Ref. \onlinecite{breitwischVLSI2007}. The phase-change material is doped Ge$_2$Sb$_2$Te$_5$. The bottom electrode has a radius of $\sim 20$ nm and a length of $\sim 65$ nm.
The phase-change material is $\sim 100$ nm thick and extends to the top electrode, whose radius is $\sim 100$ nm. Two types of devices are available on-chip that differ in the size of the
access transistor. The first sub-array contains 2 million devices, and each device is accessed by a \unit[240]{nm}-wide transistor. The second
sub-array contains 1 million devices, and two \unit[240]{nm}-wide access transistors are used in parallel per PCM element. All experiments performed in this work were done on the second sub-array, which is organized as a
matrix of 512 WL and 2048 BL.

A PCM device is selected by serially addressing a WL and a BL. To read a PCM device, the selected BL is biased to a constant voltage (typically \unit[$100-300$]{mV}) by a voltage regulator via a voltage generated off chip. The sensed current is integrated by a capacitor, and the resulting voltage is then digitized by the on-chip
8-bit cyclic analog-to-digital converter (ADC). The total duration of one read is \unit[$1$]{$\mu$s}. The readout characteristic is calibrated via on-chip reference polysilicon resistors. To program a PCM device, a voltage generated off chip is converted on chip into a programming current. This
current is then mirrored into the selected BL for the desired duration of the programming pulse. Each programming pulse is a box-type rectangular pulse with duration of \unit[400]{ns} and an amplitude varying between 0 and \unit[500]{$\mu$A}. Iterative programming involving a sequence of program-and-verify steps is used to program the PCM devices to the desired conductance values.\cite{papandreouISCAS2011} After each programming pulse, a verify step is performed, and the value of the device conductance programmed in the preceding iteration is read at a voltage of \unit[0.2]{V}. The programming current applied to the PCM device in the
subsequent iteration is adapted according to the sign of the value of the error between the target level and the read value of the device conductance. The
programming sequence ends when the error between the target conductance and the programmed conductance of the device is smaller than a margin of \unit[1.74]{$\mu$S}
or when the maximum number of iterations (20) has been reached. The total duration of one program-and-verify step is approx. \unit[$2.5$]{$\mu$s}. 

More details about the hardware platform and chip characterization results can be found in Supplementary Note I.

\subsection*{Solving the linear system for model covariance matrices.}

We solved the linear system with mixed-precision in-memory computing for the covariance matrices defined by Eq.\ \eqref{eqn:modelmatrix}. The entries of $b$ were generated uniformly in $[0,1]$. 
We used the following Conjugate Gradient (CG)
method as the inner Krylov-subspace solver:
\begin{algorithm}[H]
\caption{Conjugate Gradient with in-memory computing}\label{CG}
\begin{algorithmic}[1]
\STATE Given $r$ and initial values $z^{(1)}:=0$, $\rho^{(1)}=r$, $v^{(1)}:=r$
\FOR {$k=1,...,m$}
\STATE $w^{(k)}:=\tilde{A}v^{(k)}$ (Compute in memristive array)
%\STATE $w^k:=w^k+I_dv^k$
\STATE $\alpha^{(k)} := \langle \rho^{(k)},\rho^{(k)} \rangle/\langle w^{(k)},v^{(k)} \rangle$
\STATE $z^{(k+1)} := z^{(k)} + \alpha^{(k)} v^{(k)}$
\STATE $\rho^{(k+1)} := \rho^{(k)} - \alpha^{(k)} w^{(k)}$
\STATE $\beta^{(k)} := \langle \rho^{(k+1)},\rho^{(k+1)} \rangle/\langle \rho^{(k)},\rho^{(k)}\rangle$
\STATE $v^{(k+1)} := \rho^{(k+1)} + \beta^{(k)} v^{(k)}$
\ENDFOR
\end{algorithmic}
\end{algorithm}
The final solution is given by $z^{(m+1)}$. $\tilde{A}$ denotes the matrix which is coded in the PCM chip. When the full matrix $A$ is coded in the PCM chip, $\tilde{A} = A$. When a reduced banded version of $A$ is coded in the PCM chip, we have 
\[
\tilde{A}_{ij}=
\begin{cases}
A_{ij} ,& \text{if } 0 \leq |i-j|\leq 12 \\
0,              & \text{otherwise.}
\end{cases}
\]
All nonzero elements of $\tilde{A}$ were coded in the PCM chip using $K$ devices
averaged per element according to the procedure described in Supplementary Note III. The matrix-vector multiplication $w^{(k)}:=\tilde{A}v^{(k)}$ was performed with the chip as described in Supplementary Note III. The number of CG iterations was set to $m=5$ for $N=500$ and to $m=10$ for $N=5,000$.
The tolerance
of the iterative refinement algorithm was set to $tol = 10^{-5}$. The drift calibration procedure described in Supplementary Note V was
performed at every first iteration of Algorithm \ref{CG} on $S = 10,000$ devices to prevent errors in the multiplication results due to
conductance drift of the phase-change devices.

\subsection*{Estimation of gene interactions from RNA measurements.}

We used RNA-Seq (Illumina HiSeq 2000 RNA Sequencing Version 2) Level 3 data from TCGA (http://cancergenome.nih.gov/) project, normalized with
RSEM\cite{Y2011liBMC}. The 40 genes studied were selected from the autophagy pathway (hsa04140) curated by KEGG (http://www.genome.jp/). We
considered tissue samples across different cancer types. The number of normal samples (946) was smaller than the number of cancer samples (11935). To
compare networks estimated with the same sample size, we subsampled 946 RNA-Seq cancer profiles to match the size of the normal samples. To ensure that no bias was introduced by subsampling, we performed a Kolmogorov--Smirnov test on the subsampled distributions, which showed no evidence of difference ($p<0.05$, with Bonferroni correction). The sample covariance $A_{ij}$ between gene $i$ and gene $j$ was computed as
\begin{equation*}
A_{ij} = \frac{1}{945}\sum_{s=1}^{946}(X_{si} - \mu_i)(X_{sj} - \mu_j)
\end{equation*}
where $X_{si}$ indicates the expression value of gene $i$ in sample $s$ and $\mu_i$ is the mean expression of gene $i$ across all samples. 

The inverse
covariance matrix $\Sigma$ was computed from covariance matrix $A$ by solving $Ax^{(n)}=e^{(n)}$ for $n=1,...,40$ with mixed-precision in-memory computing. All $e^{(n)}$ entries are equal to zero
except the $n$-th one, which is 1, and $x^{(n)}$ is the resulting $n$-th column of $\Sigma$. We apply a diagonal preconditioner $M=\mathrm{diag}(A)$ on the
linear system, thus solving the problem $M^{-1}Ax^{(n)}=M^{-1}e^{(n)}$. Then, we define matrix $\tilde{A}$ as
\[
\tilde{A}_{ij}=
\begin{cases}
(M^{-1}A)_{ij} ,& \text{if } i \neq j \\
0,              & \text{otherwise.}
\end{cases}
\]
Note that all diagonal elements of $\tilde{A}$ are set to 0. All nonzero elements of $\tilde{A}$ were coded in the PCM chip using 4 devices
averaged per element according to the procedure described in Supplementary Note III. We used the following Generalized Minimum Residual
(GMRES) method as the inner Krylov-subspace solver:

\begin{algorithm}[H]
\caption{GMRES with in-memory computing}\label{GMRES}
\begin{algorithmic}[1]
\STATE Given $r$ and initial values $\beta:=\Vert r \Vert_2$, $v^{(1)}=r/\beta$
\FOR {$k=1,...,m$}
\STATE $w^{(k)}:=\tilde{A}v^{(k)}$ (Compute in memristive array)
\STATE $w^{(k)}:=w^{(k)}+I_dv^{(k)}$
\FOR {$l=1,...,k$}
\STATE $h_{lk}:=\langle w^{(k)},v^{(l)} \rangle$
\STATE $w^{(k)}:=w^{(k)}-h_{lk}v^{(l)}$
\ENDFOR
\STATE $h_{k+1,k} = \Vert w^{(k)} \Vert_2$. If $h_{k+1,k}=0$, set $m:=k$ and go to 12.
\STATE $v^{(k+1)} = w^{(k)}/h_{k+1,k}$
\ENDFOR
\STATE Define $H^{(m)} = \lbrace h_{lk} \rbrace_{1\leq l\leq m+1,1\leq k \leq m}$ and $V^{(m)}$ the matrix with column vectors $v^{(1)},..., v^{(m)}$.
\STATE Compute $y^{(m)}$ the minimizer of $\Vert \beta e^{(1)} - H^{(m)} y \Vert_2$ and $z^{(m)} = V^{(m)}y^{(m)}$.
\end{algorithmic}
\end{algorithm}

The final solution is given by $z^{(m)}$. The matrix-vector multiplication $w^{(k)}:=\tilde{A}v^{(k)}$ was performed with the chip as described in Supplementary Note
III. Line 4 of Algorithm \ref{GMRES} adds the remaining term $I_dv^{(k)}$ to $w^{(k)}$, where $I_d$ is the identity matrix (concretely, we
approximate $M^{-1}Av^{(k)}$ by $(\tilde{A} + I_d)v^{(k)}$). This avoids coding the diagonal elements of $M^{-1}A$, which are all 1s, in the memristive array to prevent unnecessary large perturbations in $w^{(k)}$ which would come from inexact computing of $I_dv^{(k)}$. The
number of GMRES iterations was set to $m=5$. The tolerance of the iterative refinement algorithm was set to $tol = 10^{-3}$. The drift calibration
procedure described in Supplementary Note V was performed at every first iteration of Algorithm \ref{GMRES} to prevent errors in the
multiplication results due to conductance drift of the phase-change devices.

The partial correlation $\rho_{ij}$ between gene $i$ and gene $j$ was computed from the inverse covariance $\Sigma$ as
\[
\rho_{ij}=
\begin{cases}
-\frac{\Sigma_{ij}}{ \sqrt{\Sigma_{ii} \Sigma_{jj} }  }, & i \neq j  \\
1,      & \text{otherwise}\\
\end{cases}
\]
For the interactome visualization, we considered interactions only for which the magnitude of the partial correlations was larger than a threshold $\tau = 0.13$, corresponding to 90th percentile of the normals. In the graphs, genes were grouped following the KEGG Orthology ($KO$) System. We defined a set of interactions between groups:
\begin{equation*}
\mathcal{I_{\alpha\beta}} = \lbrace (i,j) : i \in KO_\alpha,\, j \in KO_\beta \text{ s.t. } | \rho_{ij} |  > \tau \rbrace
\end{equation*}
where  $KO_\alpha$ and $KO_\beta$ contain gene indexes as they appear in the partial correlations matrix and correspond to the $KO$ groups defined in KEGG (e.g. AMPK or ULK).
The strength of the correlation between groups was then defined by averaging the partial correlations:
\begin{equation*}
S_{KO_\alpha, KO_\beta} =  \frac{1}{|\mathcal{I_{\alpha\beta}}|} \sum_{(i,j) \in \mathcal{I_{\alpha\beta}}} \rho_{ij}. 
\end{equation*}
Finally, the graphs were built using the partial correlations values and the strength of the correlations between groups as defined above. This allowed us to connect interacting genes and groups with a variable intensity in a more compact representation.

\subsection*{Data availability.}
The data that support the plots within this paper and other findings of this study are available from the corresponding author upon reasonable request. 
%Correspondence and requests for materials should be addressed to M.L. (anu@zurich.ibm.com) and A.S. (ase@zurich.ibm.com).

\clearpage
\section*{References}
\bibliography{papers_mem}

\section*{Acknowledgments}
We thank C. Malossi, M. Rodriguez, C. Hagleitner, L. Kull and T. Toifl for discussions; N. Papandreou, A. Athmanathan and U. Egger for experimental help; T. Delbruck for reviewing the manuscript; and
C. Bolliger for help with the preparation of the manuscript. A. S. would like to acknowledge funding from the European Research Council (ERC) under the European Union's Horizon 2020 research and innovation programme (grant agreement No. 682675).

\section*{Author contributions}
M.L., A.S., T.T., C.B., A.C. and E.E. conceived the concept of mixed-precision in-memory computing. M.L., A.S. and C.B. designed the research. M.L.
implemented the mixed-precision in-memory computing system and performed all experiments. M.L., R.M. and M.M. performed the research on the RNA expression
data. H.G. performed the evaluation of the runtime and energy consumption. All authors contributed to the analysis and interpretation of the results. M.L. and A.S. co-wrote the manuscript based on the input from all
authors.

\section*{Competing interests}
The authors declare no competing financial interests.

\end{document}